\documentclass[aps,pra,nofootinbib,preprint,amsmath,amssymb,floatfix]{revtex4}
\usepackage{graphicx}

\begin{document}

\title{The uniform quantized electron gas revisited}         % Enter your title between curly braces

\author{Enrique Lomba$^1$ and Johan S. H\o ye$^2$}        % Enter your name between curly braces
\affiliation{$^1$Instituto de Qu{\' \i}mica F{\' \i}sica Rocasolano, CSIC,
  Calle Serrano 119, E-28026 Madrid, Spain\\
$^2$Institutt for Fysikk,  NTNU, N-7491 Trondheim, Norway }

\date{\today}          % Enter your date or \today between curly braces
%\maketitle

\begin{abstract}

In this work we continue and extend our recent work on the correlation
energy of the quantized electron gas of uniform density at temperature
$T=0$. As before we utilize the methods, properties, and results
obtained by means of classical statistical mechanics. These were extended to
quantized systems via the Feynman path integral formalism. The latter
translates the quantum problem into  a classical polymer problem in
four dimensions. Again 
the well known RPA (random phase approximation) is recovered as a basic result
which we then modify and improve upon. Here we will we focus upon thermodynamic
self-consistency. Our numerical calculations  exhibit a remarkable
agreement with well known
results of a standard parametrization of Monte Carlo correlation
energies. 

\end{abstract}
\maketitle

\bigskip
\section{Introduction} 
\label{sec1}

In a recent work we considered the problem of the correlation energy
of the quantized electron gas based upon methods, results, and
reasonings of classical statistical mechanics \cite{hoye16}. With this
development the well known RPA of quantized system was recovered as a
basic result for the correlation energy. As already known, RPA results
deviate clearly from available Monte Carlo data \cite{lein00}. Since the RPA is a
basic results also from the viewpoint and methods of liquid state
theory for classical fluids, the problem will be how to utilize these
methods to improve the RPA in the quantum case.  In our previous work
we focused upon conditions 
on the pair correlation function. These must reflect the fact that
fermions with equal spins are 
not allowed to be at the same position. For classical systems the
repulsive Coulomb interaction will also prevent particles to be at the
same position. These conditions can be applied to the equal time
correlation function to improve results. The first condition is exact
while the second turns out not to be so in the quantum mechanical
case, especially for increasing energy. For this reason, in Ref.~\cite{hoye16} the second condition had
to be abandoned with some increase of uncertainty in the results. 

Our method to improve upon the RPA rests on the introduction of  an effective
interaction. In the correlation function of the RPA the pair
interaction plays a role similar to the direct correlation function
for classical systems. It is then known that for a pair of particles
at large separation ${\bf r}$, the direct correlation function
approaches the interaction times inverse temperature with minus
sign. For smaller separations there will be deviations. But they
should change smoothly for interactions that vary likewise. One then
expects the effective interaction to behave in a similar way. With
this the fermion condition  was used to determine a free parameter
in the function describing these deviations. This function may be a
Yukawa term (shielded Coulomb) or something related whose range can be
determined from the given condition. 

In the present work we will focus upon thermodynamic
self-consistency. Therefore,  the core condition with its uncertainty for
unequal spins mentioned above, will not be imposed. Self-consistency
in our system has its analog for classical systems in the SCOZA
(self-consistent Ornstein-Zernike 
approximation) and HRT (hierarchical reference theory) for classical
systems \cite{hoye77,Parola2012}. They both require thermodynamic
self-consistency between compressibility relation and internal energy
or the similar type of integral resulting from change in free energy
by a small change in pair interaction. With SCOZA the strength of the effective
interaction is tuned by changing the effective temperature, while with HRT the
effective interaction is built (and subsequently modified) by  gradually
including its wave vectors in Fourier space (renormalization). Both
SCOZA and HRT turned out to give 
very accurate results, also in the challenging critical region
\cite{Pini2002,Parola2012,lomba14}.

In the context of the electron gas correlation energy problem, in
Ref.~\cite{lein00} the ALDA (adiabatic local density 
approximation), among other approximations was considered. There the
compressibility was used to determine a parameter in an effective
interaction. However, in the ALDA formulation, this reduces to the
addition of  a constant to the Fourier
transform of the Coulomb interaction. This constant  corresponds to
adding a $\delta$-function in ${\bf r}$-space. Then the free energy is
obtained by integrating the energy-type charging integral. The result
obtained clearly over-compensates the  inaccuracy of the RPA \cite{lein00}.  

The main problem with ALDA, as we see it, lies in its use of the
$\delta$-function. A more smooth cut-off of the Coulomb interaction is
needed as mentioned above. The adequacy of this assumption was already
tested  in Ref.~\cite{hoye16}. [It may be  mentioned that for an
  interaction of finite range, the SCOZA and HRT 
  simply multiply it with an adjustable constant. This, however, is
  not suitable for the $1/r$ interaction due to its crucial tail for
  large $r\rightarrow\infty$  that should be kept  unmodified. So in this work we will investigate how
the 
use of self-consistency with a smooth cut-off can actually improve the
RPA and ALDA results, while retaining the conceptual simplicity of the
liquid-state approach explored in Ref.~\cite{hoye16}.

In Sec.~\ref{sec2} the compressibility relation for the quantized
electron gas will be established. 
Also explicit expressions for the reference system correlation
function (non-interacting Coulomb gas) and its RPA form are written
down.
  It is verified that the reference system correlation function
  when used in the compressibility relation, reproduces the known
  equation of state for free fermions. Further for interacting
  fermions the correction to the reference system equation of state
  can be expressed via the compressibility relation in terms of a
  cut-off parameter $\kappa$ in the resulting effective
  interaction. This simplifies the situation. The ALDA also evaluates
  this parameter $\kappa$ from compressibility based  known
  results. Then by computation of the correlation energy it is found
  that the inaccuracy of the RPA is over-compensated such that
  significant improvements of results are not obtained with ALDA
  \cite{lein00}. To us this inaccuracy is due to the crude
  approximation of including a constant in Fourier space or a
  $\delta$-function in ${\bf r}$-space in the effective
  interaction. To improve upon this situation we will use $\kappa$ as
  a cut-off parameter in smooth effective interactions and then study
  variations and accuracy of results.

In Sec.~\ref{sec3} the compressibility with exchange energy only,
is considered, and its influence upon the correlation function is
written down.  It is then found that the exchange energy included
  alone gives the very simple result $\kappa=\mbox{const}=1$.  

Then in Sec.~\ref{sec4} the influence of the correlation
energy is taken into account.  Since the correlation energy is a
  correction to the exchange energy, it turns out that the former has
  small influence upon the parameter $\kappa$; it decreases a small
  amount. This change will again modify the correlation energy to
  modify $\kappa$ a bit further. But this modification will be so
  small compared with other uncertainties that we simply neglect it.
  Also the precise relation of consistency between free energy and
  compressibility can be expressed through a differential equation as
  in SCOZA and HRT \cite{Pini2002,Parola2012}. But with the small
  change in $\kappa$ all this can be disregarded, and we have a much
  simplified consistency problem. With the small influence of the
  correlation energy upon the $\kappa$ we can simplify further by just
  using the correlation energy from fitted simulations to obtain
  $\kappa$.

Thus in Sec.~\ref{sec5} the ``exact'' correlation
energy, fitted from simulations, is 
expressed in terms of an analytic expression from which
compressibility  and thus the parameter $\kappa$  is
easily obtained by differentiation. 

In Sec.~\ref{sec6}
thermodynamic self-consistency between 
adiabatic charging and compressibility is considered.  Then an
  average value of $\kappa$ is used in various smooth effective
  interactions to see the accuracy by which various cutoff functions reproduce the
  simulation results.  From this the correlation function and the
wave vector dependence of contributions to the free energy are also
obtained. By including another free parameter the exact free energy is
relatively easy to recover.  Its wave vector dependence, which we
also evaluate, is a more strict test on the form of the effective
interaction of the uniform electron gas.

Our numerical results for the correlation energy (including its wave
vector dependence) are introduced in Sec.~\ref{sec7} and compared 
with the  well known Perdew-Wang parameterization \cite{lein00,PW} of
the Ceperley and Alder Monte Carlo results \cite{Ceperley1978}. We will
see how a single parameter fit of the correlation energy reproduces
its wave vector dependence over the entire range of electron
densities.  Summary and conclusions are presented  in Sec.~\ref{sec8}.

Finally  in Sec.~\ref{app} Appendix it is shown that for the Coulomb
gas the quantum mechanical virial theorem is fully consistent with the
charging principle.

\section{Compressibility relation}
\label{sec2}

The compressibility relation is valid for quantized fluids as it is
for classical fluids. So, if one considers the free fermion  gas,
the number of particles is given by Eq.~(I15) as ($T\rightarrow 0$) 
\begin{equation}
\rho=\frac{g}{(2\pi)^3}\int\frac{\zeta X}{1+\zeta X}\, d{\bf k}\rightarrow\frac{g}{(2\pi)^3}\int_{k<k_f} d{\bf k}=\frac{4\pi g}{3(2\pi)^3}k_f^3.
\label{10}
\end{equation}
Here and below the numeral I will be used to designate equations of
Ref.~\cite{hoye16}. In Eq.~(\ref{10}) the $g=2$ is the spin degeneracy
of electrons, $\zeta=e^{\beta\mu}$ where $\mu$ is the chemical
potential, and $\beta=1/(k_B T)$ where $k_B$ is Boltzmann's constant
and $T$ is temperature. Further 
\begin{equation}
X=F_\beta (k), \quad F_\lambda(k)=\exp{(-\lambda E(k))}, \quad {\rm and}
\quad E(k)=\frac{1}{2m}(\hbar k)^2        
\label{11}
\end{equation}
where $m$ is particle mass and $k$ is wave vector.

For the free fermion gas at $T=0$ the $\mu =\mu_f=(\hbar k_f)^2/2m$ where $k_f$ is the Fermi wave-vector.
By differentiation of Eq.~(\ref{10}) one finds 
\begin{equation}
\frac{\partial\rho}{\partial(\beta\mu)}=\frac{g}{(2\pi)^3}\int\frac{\zeta X}{(1+\zeta X)^2}\, d{\bf k}=g\tilde S(0,0)=\frac{g}{\beta}\hat S(0,0)
\label{12}
\end{equation}
where $\tilde S(\lambda,k)$ and $\hat S(K,k)$ are given by Eqs.~(I1) and (I4).
\begin{equation}
{\tilde S}(\lambda,k)=\frac{\zeta}{(2\pi)^3}\int\frac{{\tilde
    F}_\lambda (k'){\tilde F}_{\beta-\lambda}(k'')}{(1\pm \zeta
  X)(1\pm \zeta Y)}\,d{\bf k'} 
\label{13a}
\end{equation}
where ${\bf k''}={\bf k}-{\bf k'}$ and $\lambda=it/\hbar$ ($t$ is time), $0<\lambda<\beta$, and 
\begin{equation}
{\hat S}(K,k)=\frac{\zeta}{(2\pi)^3}\int\frac{1}{iK+\Delta}\frac{X-Y}{(1\pm \zeta X)(1\pm \zeta Y)}\,d{\bf k'}
\label{13b}
\end{equation}
here with $X=F_\beta (k')$,  $Y=F_\beta (k'')$, and
$\Delta=E(k'')-E(k')$, the plus and minus signs denoting fermions and bosons respectively.
The $S(\lambda,r)$ is the pair correlation function (including self-correlation of a single particle) in space and imaginary time $\lambda$ ($0<\lambda< \beta$). The tilde denotes Fourier transform in space with variable ${\bf k}$ while the hat also includes Fourier transform in imaginary time with variable $K$.
Then $K=2\pi n/\beta$ with $n$ integer are also the same as the well known
Matsubara frequencies. With Eqs.~(I10) - (I12) one has explicitly
\cite{lein00} 
\begin{equation}
g\hat S(K,k)=\frac{mk_f}{2\pi^2\hbar^2}f(Q,x)
\label{14}
\end{equation}
\begin{eqnarray}
\nonumber
\displaystyle
f(Q,x)&=&-\left[\frac{Q^2-x^2-1}{4Q}\ln\left(\frac{x^2+(Q+1)^2}{x^2+(Q-1)^2}\right)\right.\\
&&\left.-1+x\arctan\left(\frac{1+Q}{x}\right)+x\arctan\left(\frac{1-Q}{x}\right)\right]
\label{15}
\end{eqnarray}
with
\begin{equation}
x=\frac{mK}{\hbar^2k k_f}=\frac{K}{4\mu_f Q},\quad \mu_f=\frac{(\hbar k_f)^2}{2m}, \quad Q=\frac{k}{2 k_f}.
\label{16}
\end{equation}

 The $\beta$ can be removed from Eq.~(\ref{12}) to obtain for its inverse
\begin{equation}
\frac{\partial\mu}{\partial\rho}=\frac{1}{g\hat S(0,0)}. 
\label{17}
\end{equation} 
For interacting systems the $\hat S(K,k)$  is to be replaced by the resulting correlation function $\hat\Gamma (K,k)$. In our case it has the RPA form (I25)
\begin{equation}
g\hat\Gamma(K,k)=\frac{g\hat S(K,k)}{1+\hat A(K)},
\label{18}
\end{equation}
\begin{equation}
\hat A(K)=\hat A_e(K)=-g\hat S(K,k)(-\tilde\psi(k))=D\frac{f(Q,x)}{Q^2}, 
\label{19}
\end{equation}
\begin{eqnarray}
\nonumber
\tilde\psi(k)&=&\frac{e^2}{\varepsilon_0 k^2}, \quad \frac{4\pi}{3}(r_s a_0)^3=\frac{1}{\rho}, \quad a_0=\frac{4\pi\varepsilon_0\hbar^2}{me^2}, \\
D&=&\frac{m k_f}{2\pi^2\hbar^2}\frac{e^2}{\varepsilon_0(2k_f)^2}=0.082293\cdot r_s 
\label{20}
\end{eqnarray}
as given by Eqs.~(I5), (I9), (I13), (I16), and (I17). The
$\tilde\psi(k)$ is the Fourier transform of the Coulomb interaction,
$a_0$ is the Bohr radius, $-e$ is the electron charge, and
$\varepsilon_0$ is the vacuum permitivity.

Now the interaction can be replaced by the effective interaction (I26)
\begin{equation}
\tilde\psi_e(k)=\tilde\psi(k)L(Q), \quad L(0)=1
\label{22}
\end{equation}
by which
\begin{equation}
\hat A(K)=\hat A_e(K)=-g\hat S(K,k)(-\tilde\psi_e(k)),
\end{equation}
and relation (\ref{17}) with $\hat S$ replaced by $\hat\Gamma$ turns into
\begin{equation}
\frac{\partial\mu}{\partial\rho}=\frac{1}{g\hat \Gamma(0,0)}=\frac{1}{g\hat S(0,0)}+\tilde\psi_e(0).
\label{24}
\end{equation}
Now the $\tilde\psi_e(k)$ like the $\tilde\psi(k)$ diverges when the
$k\rightarrow0$. However, the electron gas is neutralized by a
neutralizing background that cancels the mean field term
$\tilde\psi(0)$ (small charges $\rightarrow0$ of high density
$\rightarrow\infty$). Thus with pressure $p$ we end up with the
compressibility relation ($Q\rightarrow 0$)
\begin{eqnarray}
\nonumber
\frac{1}{\rho}\frac{\partial p}{\partial\rho}&=&\frac{\partial\mu}{\partial\rho}=\frac{1}{g\hat S(0,0)}+(\tilde\psi_e(Q)-\tilde\psi(Q)) \quad\quad \\
&=&\frac{1}{g\hat S(0,0)}+(L(Q)-1)\tilde\psi(Q)=\frac{2\pi^2\hbar^2}{m k_f f(0,0)}-\frac{1}{\kappa^2}\frac{e^2}{\varepsilon_0(2k_f)^2}
\label{25}
\end{eqnarray}
with $f(Q,x)$ and $\tilde\psi(k)$ given by Eqs.~(\ref{15}) and (\ref{20}) respectively.
For small $Q$ we have used the expansion 
\begin{equation}
L(Q)=1-\frac{Q^2}{\kappa^2}.
\label{26}
\end{equation}
Thus compressibility determines the coefficient $\kappa$.

\section{Compressibility with exchange energy} 
\label{sec3}

The chemical potential of the free electron gas is the well known ($T=0$)
\begin{equation}
\mu_0=\mu_f=\frac{(\hbar k_f)^2}{2m}\sim k_f^2\sim\rho^{2/3}
\label{30}
\end{equation}
with $\rho$ given by  (\ref{10}). Helmholtz free energy per unit volume is then 
\begin{equation}
F_0=\frac{3}{5}\rho\mu_f\sim\rho^{5/3} 
\label{31}
\end{equation}
since $\mu_0=\partial F_0/\partial\rho$. This gives the inverse compressibility (divided by $\rho$)
\begin{equation}
\frac{\partial\mu_0}{\partial\rho}=\frac{\partial^2 F_0}{\partial\rho^2}=\frac{2\mu_f}{3\rho}=\frac{\pi^2\hbar^2}{m k_f}.
\label{32}
\end{equation}
This is precisely the first term of relation (\ref{25}) as should be expected noting that from expression (\ref{15}) the $f(0,0)=2$.

The exchange energy per unit volume $F_{ex}$ is given by Eqs.~(9.5) and (9.9) of Ref.~\cite{hoye11} as
\begin{equation}
F_{ex}=\frac{g}{2(2\pi)^3}\int\tilde S(0,k) \tilde\psi(k)\,d{\bf k}=-\frac{3\pi e^2 k_f}{2(2\pi)^3 \varepsilon_0}\rho\sim \rho^{4/3}.
\label{33}
\end{equation}
The contribution to the inverse compressibility thus becomes
\begin{equation}
\frac{\partial^2 F_{ex}}{\partial \rho^2}=\frac{1}{3}\cdot\frac{4}{3}\cdot\frac{F_{ex}}{\rho^2}=-\frac{e^2}{\varepsilon_0(2k_f)^2}.
\label{34}
\end{equation}
When comparing with the last term of Eq.~(\ref{25}) one finds that the exchange energy alone means that $\kappa=1$.

The exchange energy is the dominating contribution to the interaction
energy at high density. Thus $\kappa=1$ is the high density value
while for finite densities there will be higher order corrections to
this value. Altogether we will find that $\kappa$ has small variations
consistent with numerical results obtained in Ref.~\cite{hoye16}. 

\section{Use of ``exact'' correlation energy} 
\label{sec4}

For the electron gas of uniform density accurate results for the
correlation energy have been obtained from simulations \cite{Ceperley1978}. Thus these
results which may be considered exact, can be utilized to obtain
values for the parameter $\kappa$ in a rather straightforward way. For
general use it is preferable to express the correlation Helmholtz 
energy per particle $f_c$ as 
\begin{equation}
f_c=\mu_f G(r_s).
\label{40}
\end{equation}
Inserting numerical values with use of Eqs.~(\ref{10}), (\ref{20}), and (\ref{30}) one finds Eq.~(I20) for the Fermi energy
\begin{equation}
\mu_f=\frac{50.1\,\mbox{eV}}{r_s^2}.
\label{41}
\end{equation}

We will need first and second derivatives of $G(r_s)$, so it will be
advantageous to fit it with an analytic expression. Here it should be
noted that if the Coulomb interaction is scaled with a
parameter $\lambda$ ($0\leq\lambda\leq1$), i.e. $e^2\rightarrow\lambda
e^2$, the function $G(r_s)$ will scale to become $G(x)$ with
$x=\lambda r_s$ while expression (\ref{41}) for $\mu_f$ is
unchanged. This is the charging principle which we will consider
\cite{lein00}. [Note, here the parameter $\lambda$ is not the imaginary time of Eqs.~(\ref{11}) and (\ref{13a}).]
The scaling property, consistent with one free
parameter for the electron gas, can be seen by considering the
Schr\"{o}dinger equation for a particle in the Coulomb potential or
its extension to more particles with Coulomb interactions. 

With correlation energy $F_c=\rho f_c$ per unit volume we now with expressions (\ref{40}) and (\ref{41}) find for its contributions to the chemical potential and inverse compressibility ($r_s\sim1/\rho^{1/3}$)
\begin{equation}
\mu_c=\frac{\partial F_c}{\partial \rho}=\frac{\mu_f}{3}(5G-xG')
\label{42}
\end{equation}
\begin{equation}
\frac{\partial^2 F_c}{\partial \rho^2}=\frac{\mu_f}{9\rho}(10G-6xG'+x^2G'')
\label{43}
\end{equation}
where $G=G(x)$ and $G'$ and $G''$ are its derivatives with respect to
$x=\lambda r_s$. Here the charging parameter $\lambda$ (not present in
$\mu_f$) has been included. 

Expression (\ref{43}) can be directly related to expression (\ref{25})
to determine the parameter $\kappa$. With use of Eqs.~(\ref{10}), (\ref{16}), (\ref{20}),
 and (\ref{32}) expression (\ref{25}) can be rewritten in the
convenient form ($f(0,0)=2$)
\begin{equation}
\frac{\partial^2 F}{\partial \rho^2}=\frac{\partial\mu}{\partial\rho}=\frac{\mu_f}{\rho}\left(\frac{2}{3}-\frac{1}{\kappa^2}\frac{4 Dx}{3r_s}\right)
\label{44}
\end{equation}
where the parameter $\lambda$ also is included with $x=\lambda r_s$. The $F$ is the total free energy per unit volume. As already concluded, the exchange energy alone results in $\kappa=1$. Thus for the correlation energy one is left with
\begin{equation}
\frac{\partial^2 F_c}{\partial \rho^2}=-\frac{\mu_f}{\rho}H(x), \quad H(x)=\frac{4}{3}\left(\frac{1}{\kappa^2}-1\right)\frac{Dx}{r_s}.
\label{45}
\end{equation}
So we find
\begin{equation}
\frac{1}{\kappa^2}=1+\frac{3r_s}{4D}\frac{H(x)}{x} \quad \left(\frac{D}{r_s}=0.082293\right)
\label{46}
\end{equation}
where with Eqs.~(\ref{43}) and (\ref{45})
\begin{equation}
H(x)=-\frac{1}{9}\left(10G-6xG'+x^2 G''\right).
\label{47}
\end{equation}

\section{Analytic expression for exact correlation energy}
\label{sec5}

In the high density limit, $k_f\rightarrow\infty$ or $r_s\rightarrow 0$,
the exact  correlation energy is known to have the form
$f_c\sim\ln{r_s}+\mbox{const}$. Inserted in Eqs.~(\ref{40}) and
(\ref{41}) this means 
\begin{equation}
G(x)=ax^2\ln x+b
\label{50}
\end{equation}
where $a$ and $b$ are constants. Here $x=\lambda r_s$ (with
$\lambda=1$). For $r_s\rightarrow 0$ one finds a good fit to
the PW parameterization with $a=0.016$, and $b=-0.0293$.  This is near the exact limiting result \cite{limit} ($f_c=-13.6\,{\rm eV}\cdot 0.0622\cdot\ln
(a_0k_f) + {\rm const.}= -0.846\,{\rm eV}\cdot\ln
(a_0k_f) + {\rm const.}$), which gives $a=0.0169$.  In a general
situation, one will find that (\ref{50}) with a decaying  coefficient $a=a(x)$ will fit the energy well (for $r_s<10$). Thus $a$ may be replaced by the function
\begin{equation}
a(x)=a'+a_1\left(\frac{1}{1+\alpha_1 x}\right)+a_2(e^{-\alpha_2 x}-1) 
\label{51}
\end{equation}
where the coefficients $a'$, $b$, $a_1$, $\alpha_1$, $a_2$, and
$\alpha_2$ are to be determined. Here we have chosen to fit
the coefficients to reproduce the PW correlation energy over the range
$r_s=0.001$ to $r_s=10$, which gives $a'=0.0086$, $b=-0.0324$,
$a_1=0.0134$, $\alpha_1=1.002$, $a_2=-0.0067$, and $\alpha_2=0.843$. 
With this expression derivatives of
the resulting $G(x)=a(x)\,x^2(\ln x +b)$ are obtained in a
straightforward way, and one can determine the dependence of
$\kappa$ on $x$ (or $r_s$) from Eq.~(\ref{46}). 

\begin{figure}[h!]
\centering
\includegraphics*[scale=0.7,angle=0,clip]{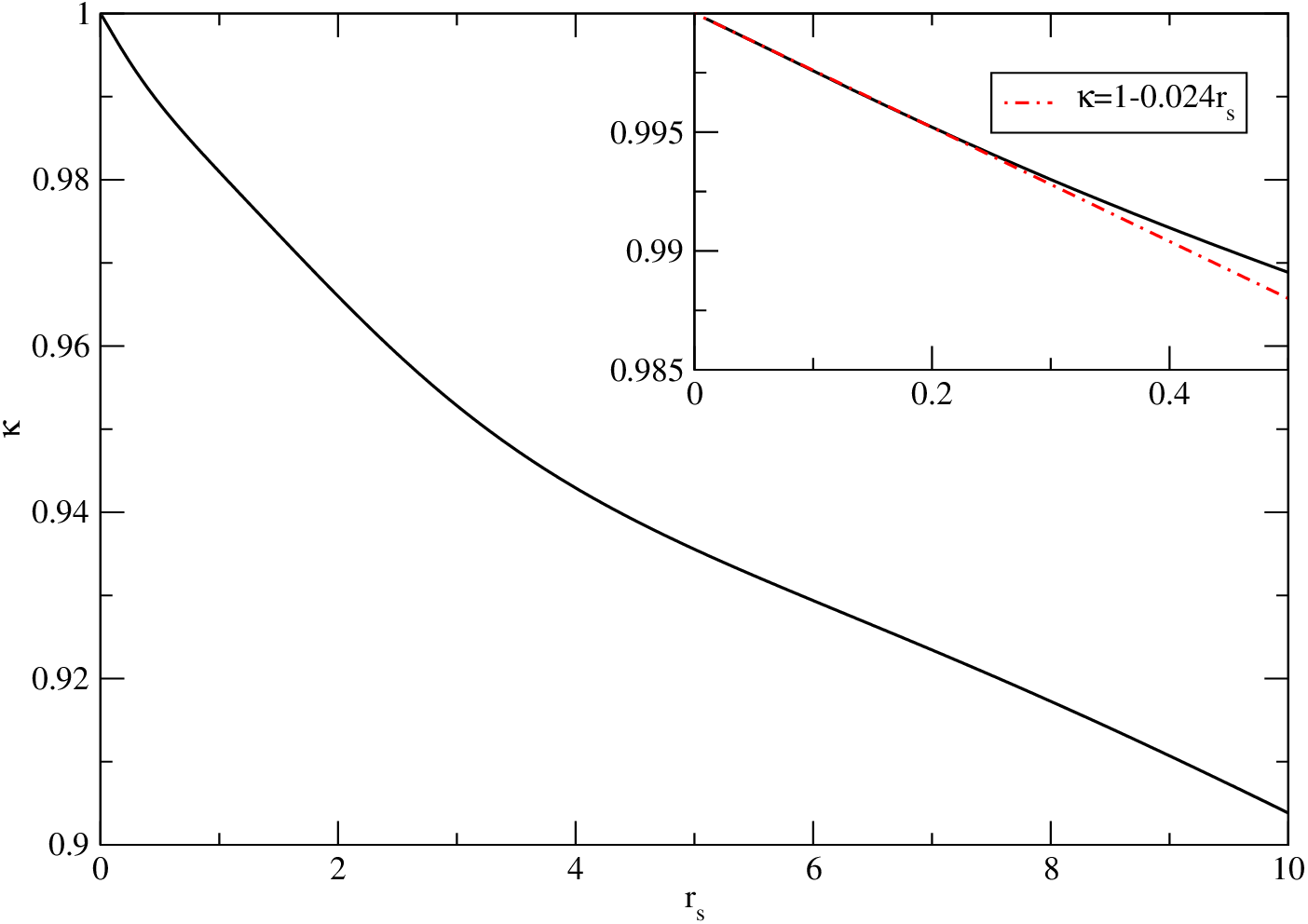}
\centering
\caption{Dependence of  the $\kappa$  parameter -- see Eqs.(\ref{25}) and
  (\ref{46}) -- on the $r_s$. The inset illustrates the high electron
  density regime, with the dash-dotted line corresponding to the
  asymptotic behavior given by Eq.(\ref{53}).}
\label{kappa}
\end{figure}

Also, using expression (\ref{50}) the small $x$ behavior of the coefficient
$\kappa$ of the cutoff function (\ref{26}) is easily obtained. Thus by
differentiation and insertion into Eqs.~(\ref{46}) and (\ref{47}) with
$a=0.016$ (from the $r_s\rightarrow 0$ fit) one finds 
\begin{equation}
\frac{1}{\kappa^2}\approx 1+\frac{3r_s}{4D}\frac{a}{3}x=1+3.04 ax=1+0.048 x
\label{52}
\end{equation}
\begin{equation}
\kappa\approx 1-0.024x.
\label{53}
\end{equation}
In Fig.~\ref{kappa} we present the variation of $\kappa$ as a
function of $x$, and in the inset the high density linear limit
(\ref{53}) is illustrated. Thus $\kappa$ varies little and stays near
1 (for $\lambda 
r_s=x<10$). Due to this small variation, we in our numerical work as a
great simplification have assumed $\kappa$ constant, keeping an average
value ($\kappa = 0.96$) when evaluating the correlation energy. 

\section{Thermodynamic self-consistency}
\label{sec6}

In Ref.~\cite{hoye16} we used the RPA type free energy with effective
interaction. It was inspired by the MSA (mean spherical approximation)
free energy for classical hard spheres with perturbing interaction
outside. In Ref.~\cite{lein00} its Eq.~(23), the adiabatic
approximation, is used along with the compressibility given by its Eq.~(31) to
obtain the ALDA (adiabatic local-density approximation). But there the
use of the compressibility only adds a constant to the Fourier
transform of the interaction $\tilde\psi(k)$ to obtain the effective
interaction, $\tilde\psi_e(k)=\tilde\psi(k)L(Q)$, with $L(Q)$ given by
Eq.~(\ref{26}) for all $Q$. 
This means that the added piece is merely a $\delta$-function in ${\bf
  r}$-space. Compared with results for classical fluids and the
"classical polymer problem" formed by the quantized electron gas, this
is clearly inaccurate. As argued in Ref.~\cite{hoye16}, the Coulomb
interaction 
should be cut by a smooth function to obtain the effective one where
we will assume $L(Q)\rightarrow 0$ as $Q\rightarrow\infty$. 

Based on the exact free energy the ALDA curve for $r_s=4$ is obtained
in Fig.~2 of Ref.~\cite{lein00}. [This figure gives the $Q$-dependent 
contributions $\varepsilon_c(Q)$ to the correlation energy according to 
expressions (\ref{61}) and (\ref{79}) below.]
There accurate results are  obtained
for $Q<0.7$, but they become rather inaccurate for larger values of
$Q$. Clearly this is related to the wrong large $Q$-behavior of
$L(Q)$.  We can now impose thermodynamic self-consistency using a
smooth cut-off function with a free (or an extra free) parameter to be
determined. In this way the equation of state is required to be the
same both via the free energy expression and the
compressibility. This, as mentioned earlier, corresponds to the SCOZA
and HRT methods used for classical fluids to obtain very accurate
results near the critical region\cite{Pini2002,Parola2012}. These
approaches lead to partial differential 
equations that are highly non-trivial to solve numerically. For the
electron 
gas, however, the two variables of interest enter  both through the scaling
variable $x=\lambda r_s$; so it all simplifies.  However, as mentioned
in Sec.~\ref{sec1} we can simplify further. Thus we will utilize
the known exact solution. Also the parameter $\kappa=\kappa(x)$ may
be considered constant with respect to $\lambda$, since as we have
seen, it changes little with electron density. This  leads to a
further  simplification.

The adiabatic charging principle, Eq.~(23) of Ref.~\cite{lein00},
gives the change to the Helmholtz free energy by a small change in the
perturbing potential. We first consider the RPA with correlation
energy (I18) and (I19) modified by the charging parameter $\lambda$. 
\begin{equation}
f_c=\frac{\Delta F_c}{\rho}=12\int_0^\infty\tilde f_c(k)Q^2\,dQ
\label{61}
\end{equation}
\begin{equation}
\tilde f_c(k)=\frac{1}{\pi}\mu Q\int[\ln(1+\lambda\hat A(K))-\lambda\hat A(K)]\,dx
\label{62}
\end{equation}
with  $\hat A(K)$ given by Eq.~(\ref{19}). From this we find
\begin{equation}
u_c(k)=\lambda\frac{\partial\tilde f_c(k)}{\partial\lambda}=\frac{1}{\pi}\mu Q\int\left[\frac{\lambda\hat A(K)}{1+\lambda\hat A(K)}-\lambda\hat A(K)\right]\,dx
\label{63}
\end{equation}
In terms of the correlation function this can be written as
\begin{equation}
u_c(k)=\frac{1}{\pi}\mu Q\int g[\hat\Gamma(K,k)-\hat S(K,k)]\lambda\psi(k)\,dx.
\label{64}
\end{equation}
Now this equation is not tied to the RPA, but is valid more
generally. So the problem left is to have a best possible
approximation to the resulting correlation function. Then again we
keep the MSA form (I38) (with $\hat\Gamma_+\rightarrow\hat\Gamma$,
$\hat A_+\rightarrow \hat A_e$) or Eq.~(\ref{18}) with $\hat
A\rightarrow\lambda A_e$ 
\begin{equation}
\hat\Gamma(K,k)=\frac{\hat S(K,k)}{1+\lambda\hat A_e(K)}
\label{65}
\end{equation}
where with effective interaction $\tilde\psi_e(k)=\tilde\psi(k)L(Q)$
\begin{equation}
\hat A_e(K)=\hat A(K)L(Q)
\label{66}
\end{equation}
with $\hat A(K)$ given by Eq.~(\ref{19}). It can be noted that with
expression (\ref{64}) only the $\hat\Gamma_+\rightarrow\hat\Gamma$
part contributes since $\tilde\psi(k)$ at the end of Eq.~(\ref{64}) is
the un-cut interaction. In Ref.~\cite{hoye16} a $\hat\Gamma_-$ part
was also present, but now it will not contribute
in (\ref{64}) since the un-cut interaction there at the end of the 
equation does not distinguish  between pairs of equal and unequal spins.

From the results of Ref.~\cite{hoye16} it was found that exact results
were well approximated with $\kappa$ constant. This is consistent with
the small $x$ result (\ref{53}). With this approximation expression
(\ref{64}) is easily integrated to obtain 
\begin{equation}
\tilde f_c(k)=\frac{\mu Q}{\pi L(Q)}\int[\ln(1+\lambda\hat A_e(K))-\lambda\hat A_e(K)]\,dx
\label{67}
\end{equation}
This simplifies considerably as numerical integration of (\ref{64}) is
avoided. Then an average value of $\kappa=\kappa(x)$ in the interval
$0<x<r_s$ may be used. We have extended the approximation further by keeping
$\kappa$ constant in the whole interval $0<r_s<10$. Thus the ALDA
curve of Fig.~2 in Ref.~\cite{lein00} is recovered from expressions
(\ref{26}) and (\ref{67}) with $\kappa\approx0.96$. Note this
curve changes sign at $Q=\kappa$. The reason is that for small $L(Q)$
the integrand of (\ref{67}) is proportional to $L(Q)^2$. 

One way to utilize result (\ref{67}) is to use smooth cutoff
functions like the ones used in Ref.~\cite{hoye16} or something
similar from which to evaluate the free energy that follows from
expression (\ref{67}) with $\lambda=1$. The average $\kappa$ to be
used, with $L(Q)$ expanded like Eq.~(\ref{26})), follows from
Eq.~(\ref{46}). The resulting free energy will deviate somewhat from
the exact one, but usually much less than the ALDA one, depending upon
the $L(Q)$ chosen. Also the result may be used to obtain an iterated
value for $\kappa$ to make it self-consistent. But the change will be
small anyway, and has been neglected here. 

Next one can search for a smooth cut-off function $L(Q)$ that
reproduces the exact result for the correlation energy $F_c$. This can be done by combining two
functions $L_1(Q)$ and $L_2(Q)$ with an additional free parameter
$\alpha$ 
\begin{equation}
L(Q)=\alpha L_1(Q)+(1-\alpha)L_2(Q)
\label{68}
\end{equation}
with $\kappa$ as before. Then self-consistency is imposed by requiring that
via Eqs.~(\ref{63})-(\ref{66}) integral (\ref{61}) for the free energy is the exact
one. Remaining inaccuracy then lies in the curve for the function
$f_c(k)$ where deviations of one sign are compensated by those of
opposite sign. Further the $L_2(Q)$ can be made as an additional combination of functions with free parameters as in Eq.~(\ref{lmix}) below to minimize the inaccuracy of $\tilde f_c(k)$.

It should again be noted that the self-consistency problem discussed
above is the analog of the SCOZA and HRT that both have resulted in
very accurate results for classical fluids. They both have the
internal energy type integral (\ref{64}) that computes the response to
a change in interaction, i.e.~change in inverse temperature $\beta$
and change in wave vector $Q$ respectively. Free energy then follows
by integration, and it is forced to be consistent with
compressibility. This gives rise to a non-linear partial differential
equation. In the present case, however, the situation simplifies much
due to the scaling form (\ref{40}) by which derivatives with respect
to $\lambda$ and $\rho$ are essentially the same as they join in the
scaling variable $x=\lambda r_s$. The detailed differential equation
of thermodynamic self-consistency is then obtained by equating the
$\partial^2(\lambda\partial F_c/\partial\lambda)/\partial\rho^2$ that
follows from Eqs.~(\ref{61}) - (\ref{64}) with the
$\lambda\partial(\partial^2F_c/\partial\rho^2)/\partial\lambda$ that
follows from (\ref{45}). But in view of the discussion and
simplifications made above resulting in Eq.~(\ref{67}) and use of
available exact energy, we find no reason to go further in this direction.

\section{Results}
\label{sec7}
\begin{figure}[h!]
\centering
\includegraphics*[scale=0.8,angle=0,clip]{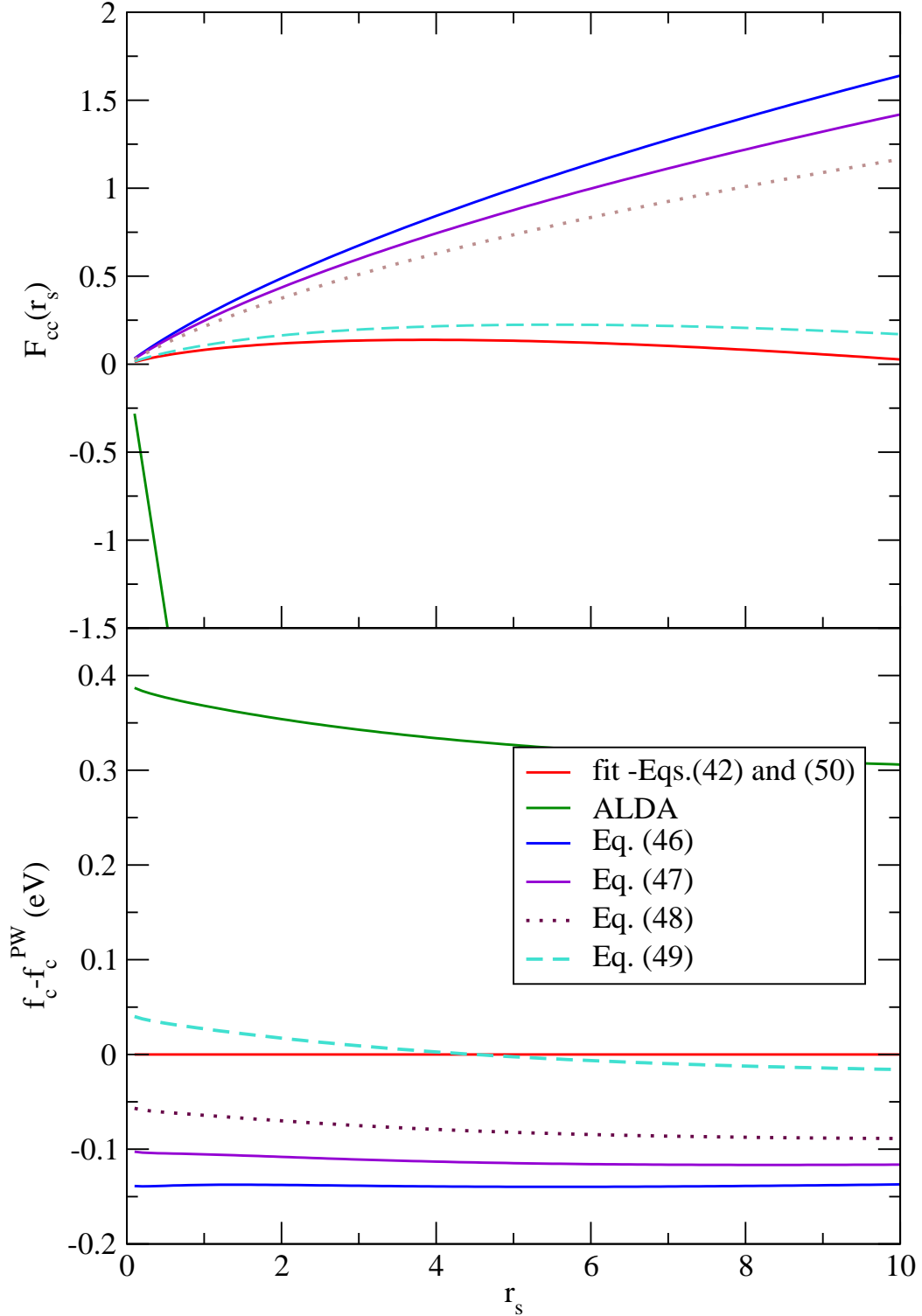}
\centering
\caption{Computation of the core condition integral Eq.~(\ref{core}) 
for various cuts (top
  graph). The  electron correlation energy with respect to
  the Perdew-Wang parameterization\cite{PW} (bottom graph) is
  computed with the
  various cuts indicated by the equation numbers and in the ALDA \cite{lein00}).}  
\label{frs}
\end{figure}

Along the same lines of Ref.~\cite{hoye16}, here we utilize
Eqs.~(\ref{61}), (\ref{66}) and (\ref{67}) with a series of cutoff
functions which are enumerated below,
\begin{eqnarray}
  \label{cut}
L_{ALDA}(Q)&=& 1-Q^2/\kappa^2\label{alda}\\
\label{cut1}
L_{se}(Q)&=&\frac{\kappa^2}{Q^2+\kappa^2},\label{se}\\
L_{exp}(Q)&=&\left(\frac{\kappa^2}{Q^2+\kappa^2}\right)^2,\label{exp}\\
L_{erf}(Q)&=&\exp{(-Q^2/(2\kappa)^2)},\label{erf}\\
\label{cut4}
L_{gauss}(Q)&=&1-Q^2\left(\left[\frac{3}{2}-\frac{Q^2}{4\kappa^2}\right]\frac{D_+(Q/(2\kappa))}{\kappa
  Q}+\frac{1}{4\kappa^2}\right),
\label{LQ}
\end{eqnarray}
with $D_+(x)$ being  Dawson's
integral \cite{hoye16,abramowitz}. As mentioned, the parameter $\kappa$ will
be kept fixed and equal to 0.96 for all calculations. Thus with these cutoffs the resulting free energies are found and compared with the PW parameterization \cite{PW} in Fig.~\ref{frs}. Additionally, the cutoff function (\ref{68}) is implemented by fitting the $\alpha$
parameter in this equation so that $f_c$ agrees with the PW
parameterization. After various trials, we have found that an optimum
fit is obtained with, $L_1(Q)=L_{gauss}(Q,\kappa)$ and
\begin{equation}
  L_2(Q) = L_{exp}(Q,\xi\kappa) -
  L_{erf}(Q,\xi\kappa)-\alpha'\left(L_{exp}(Q,\xi'\kappa) -
  L_{erf}(Q,\xi'\kappa)\right)
  \label{lmix}
\end{equation}
which combines two possible different ranges in $Q$. The procedure is
then the following, with a given set of initial parameters we use a
Marquardt-Levenberg residual minimization algorithm to fit $f_c(Q)$ and
$f_c$ to the PW value, for $r_s=4$, obtaining a set of
$(\alpha,\xi,\xi',\alpha')$ parameters. Then, for different values of
$r_s$ only the value of $\alpha'$ is adjusted to match the integrated
correlation energy $f_{c}$. In the lower graph of figure \ref{frs} we
represent the deviations of the correlation energy with respect to the
PW parameterization, for the fitted case (which obviously agrees
completely by its construction), for the ALDA, and for results from Eq.~(\ref{67})
using the various smooth cuts (\ref{cut1}) - (\ref{cut4}) for the function $L(Q)$, with $\kappa=0.96$
fixed. As found in Ref.~\cite{hoye16}, the Gaussian cut (\ref{cut4}) performs
substantially better than other cuts, clearly outperforming both the
RPA and the ALDA approximation.
 Note the ALDA curve deviates a bit from the one of Fig.~5 of Ref.~\cite{hoye16}
which was further taken from Ref.~\cite{lein00}. This deviation we 
attribute to our very simplified $\kappa=\mbox{const}$ integration with
respect to $\lambda$ which leads to expression (\ref{67}).

We can now check to what extent the  average core condition on the 
correlation function behaves due to the
approximations we have made. 
In Ref.~\cite{hoye16} it was used as a condition. However, it was realized
that particles with unequal spins and with Coulomb repulsion could, depending upon the particle density, be on the same position.
Thus only a crude estimate or guess of the core condition was made. 
From Eq.~(57) of Ref.~\cite{hoye16}, the average contribution to the core
condition on the correlation function (\ref{18}) for equal and unequal 
spin pairs can be written
\begin{equation}
F_{cc}(r_s)=\frac{36}{\pi} \int\int \frac{D f^2(Q,x)L(Q,\kappa)}{Q^2 +D
  f(Q,x)L(Q,\kappa)}dx\, Q^3\,dQ.    
\label{core}
\end{equation}
Thus, $F_{cc}$ does not contain the contribution from the reference 
system $\hat S(K,k)$. Furthermore it is here normalized such that it
will be equal to 1, instead of -1, when pairs of unequal spins cannot
overlap, and is equal to 0 when they can overlap fully.
The function $F_{cc}$ is plotted in the upper graph of Fig.~\ref{frs} 
for the same range of electron densities. Here one observes a reasonable
behavior by all approximations except for ALDA that goes rapidly negative.
But they all approach 0 when $r_s\rightarrow0$ (high density and high 
energy limit). However, for $r_s$ large (low density) 3 of the curves 
cross the hard core line located at 1. [If the exact curve had been known, 
another parameter could have been fitted.]

\begin{figure}[h!]
\centering
\includegraphics*[scale=0.7,angle=0,clip]{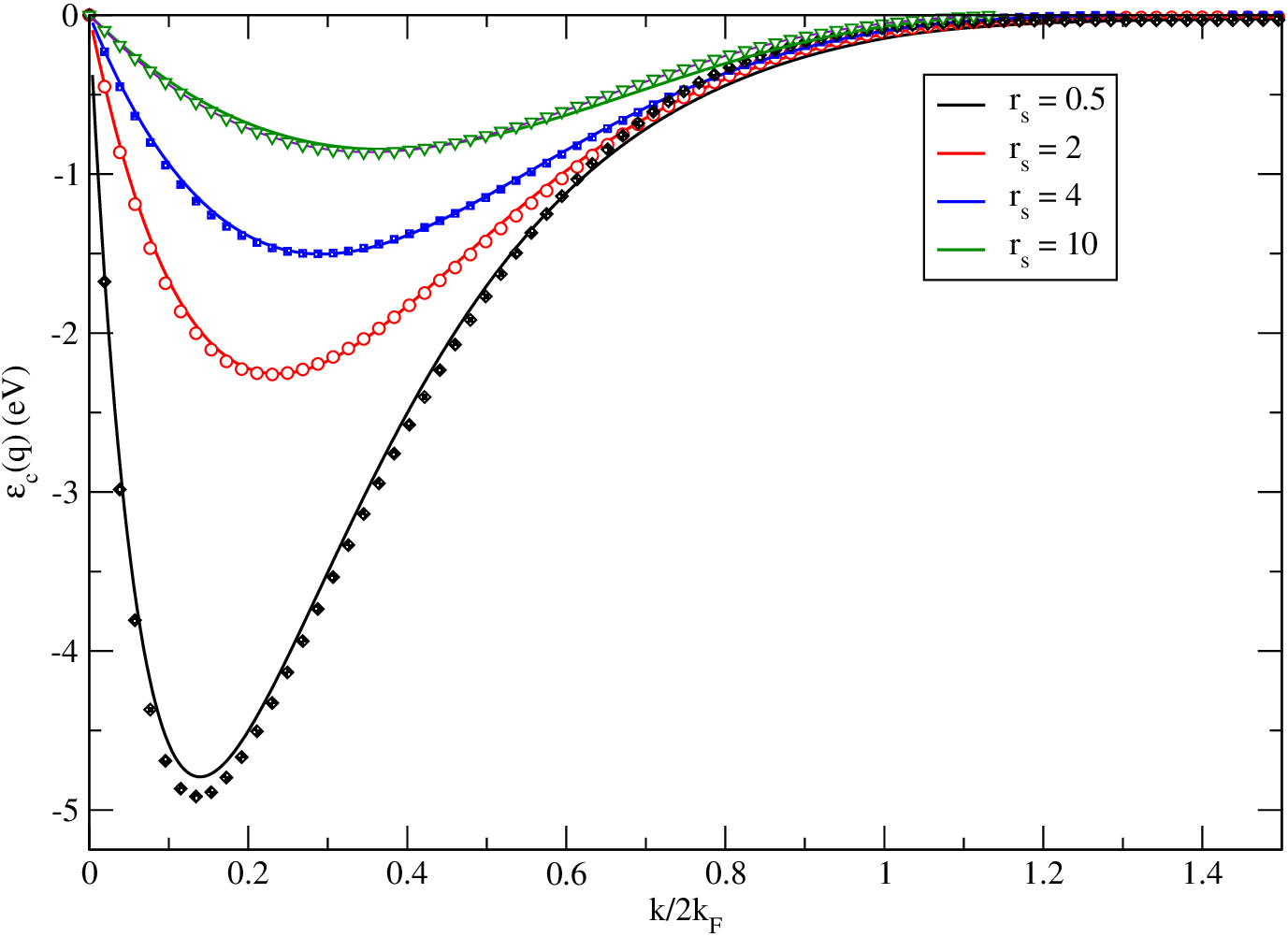}
\centering
\caption{Wave vector analysis of the correlation energy per electron,
  calculated using the optimized cut given by Eqs.~(\ref{68}) and (\ref{80}) for the
  effective interaction and then compared  with the
  Perdew-Wang parameterization \cite{lein00,PW} computed for various
  values of $r_s$.}
\label{ecorr}
\end{figure}

Finally, in Figure \ref{ecorr} we plot the wave vector decomposition
of the correlation energy for various values of $r_s$ and compare it
with those of the PW parameterization.  Then the $L(Q)$ of Eq.~(\ref{68})
together with expression (\ref{lmix}) is used.  As it is
customary we represent the function,
\begin{equation}
\varepsilon_c(Q) = 12 Q^2 \tilde f_c(Q)
\label{79}
\end{equation}
which can be directly integrated over $Q$ to yield the correlation
energy $f_c$. The results for $r_s=4$ obviously match those of the
parameterization  best --this is the fitted value for all 4 
parameters mentioned above--, but small departures
are visible for low and high electron densities. Recall that aside
from  the case $r_s=4$ only a single parameter is fitted to 
the integrated
correlation energy. The curves tend to overestimate slightly the long wave
vector contribution to the correlation, which is compensated by an
underestimation at small $k$.

\section{Summary and conclusions}
\label{sec8}

We have evaluated the correlation energy for the uniform quantized
fermion gas of electrons, extending our recent work\cite{hoye16}. In
Ref.~\onlinecite{hoye16} the core   
condition, which implies that fermions with equal spins can not be at the same
position, played a central role in our approximation. However, with
quantum mechanics particles with opposite 
spins have the possibility to be at the same position also with
Coulomb repulsion. Thus the average core condition became somewhat
undetermined. The basis for the computations were the RPA, but with a
smooth cut of the Coulomb interaction for small distances. For the
free energy an expression inspired by the MSA of classical fluids was
used. Compared with simulation data, good results were obtained for
certain cutoff functions while others led to divergence problems. 

Inspired by the ALDA of Ref.~\cite{lein00} we in this work have
focused upon the compressibility relation and thermodynamic
self-consistency as used in the SCOZA and HRT for classical fluids
\cite{hoye77,Parola2012,Pini2002,lomba14}. This determines a parameter
$\kappa$ in the effective interaction. In addition, we make use of the
fact that the
effective (cut) interaction behaves as a smooth function. The range of
the cut is the inverse of $\kappa$. With reasonable cutoff functions,
results for the correlation energy substantially improve. Also with Eqs.~(\ref{68}) and
(\ref{lmix}) one has a cut with additional free parameters to
reproduce the exact correlation energy. Finally, as a more sensitive
test of our approach for the free energy, in Fig.~\ref{ecorr} we
present its  wave vector decomposition. Also
for this more detailed analysis we find a good agreement with the simulation
results, which again reflects the proper form of the effective (cut)
interaction used. 

Our results may be extended to the non-uniform electron gas of
molecules. The basis for the correlation energy will still be the RPA
extended to this more general situation. Numerical evaluations will
clearly be much more demanding as translational symmetry is lost. The
reference system correlation function is now determined by the
eigenstates, including the excited ones, that follows from the DFT
(density functional theory) or Hartree-Fock solution of the problem
considered from which the standard RPA established. Then the RPA is
modified such that the full Coulomb interaction is replaced by a cut
one consistent with or similar to the one obtained for the uniform
case studied in this work. A crucial result of our work is that the
cut function $L(Q)$ is a smooth one that has small or negligible
variation with changing electron density. But there will be some
complication as clear density dependence sits in the shift of variable
$k=2k_f \kappa Q$ ($k_f\sim \rho^{1/3}$). Further in the non-uniform
case one needs the effective interaction in ${\bf r}$ space,
$\psi_e(r)=\psi(r) f(r)$ ($\psi(r)\sim 1/r$). So with cuts
(\ref{cut1}) - (\ref{cut4}) one has $f(r)=H(x)$ with $x=2 k_f \kappa
r$ where $H(x)$ is given by Eq.~(48) of Ref.~\cite{hoye16}. Since
${\bf r}={\bf r}_2-{\bf r}_1$ the $k_f$ ($\sim \rho^{1/3})$ to choose
to determine $\kappa$ will not be obvious. As a compromise,
some average density for positions 1 and 2 can be used. Further
investigations will need  numerical ab 
initio calculations on non-uniform system with basis on the RPA, and
are beyond the scope of this work.

\acknowledgments
EL  acknowledges the support from the Direcci\'on
General de Investigaci\'on Cient\'{\i}fica  y T\'ecnica under Grant
No. FIS2013-47350-C5-4-R.

\section{Appendix: Virial theorem}
\label{app}

Quantized systems also fulfill the virial theorem. For the
quantized electron gas we will show that it is fully consistent with
the adiabatic approximation. On the other hand this consistency means
that there will be no extra condition to determine some extra
parameter. 
The virial theorem is
\begin{equation}
p=\frac{2}{3}\rho u_k-\frac{1}{6}\int d{\bf r}\, {\bf r}\nabla\psi(r) n(r) 
\label{80}
\end{equation}
where  $p$ is pressure, $u_k$ is kinetic energy per particle, and
  $n(r)$ is the pair distribution function. This theorem is also valid
  for quantized systems. [For hard core particles one has to take the
    proper limit of the integral. This limit is different in the
    classical and quantum cases.] For an ionic fluid that is
neutralized by an oppositely charged background the $n(r)$ is replaced
by $\rho^2 h(r)$ where $h(r)$ is the pair correlation function. 

With Coulomb interaction $\psi\propto1/r$ by which
\begin{equation}
{\bf r}\nabla\psi(r)=r\frac{\partial}{\partial r}\psi(r)=-\psi(r).
\label{81}
\end{equation}
The integral in Eq.~(\ref{80}) can also be written as a Fourier integral. Then ${\bf r}\nabla\psi(r)\rightarrow -\nabla({\bf k}\tilde\psi(k))$. So with $\tilde\psi(k)\propto 1/k^2$ 
\begin{equation}
\nabla({\bf k}\tilde\psi(k))=\frac{\partial}{\partial k_i}\left(k_i \tilde\psi(k)\right)=\tilde\psi(k)
\label{82}
\end{equation}
which is equivalent to to Eq.~(\ref{81}) ($k_i k_i=k^2$, $\sum_i$, $i=1,2,3$). So with Fourier transform and use of Eq.~(\ref{82}) for Coulomb interaction the virial theorem becomes
\begin{equation}
p=\frac{2}{3}\rho u_k+\frac{1}{3}\rho u_p 
\label{83}
\end{equation}
\begin{equation}
\rho u_p=\frac{1}{2(2\pi)^3}\int d{\bf k}\,\rho^2 \tilde h(k) \tilde\psi(k).  
\label{83a}
\end{equation}
 where $u_p$ is the potential energy per particle for the neutral ionic fluid. Now the total energy per particle is the same as the internal energy $u=u_k+u_p$. With this $u_k=u-u_p$ by which
\begin{equation}
p=\frac{2}{3}\rho u-\frac{1}{3}\rho u_p. 
\label{83b}
\end{equation}

At $T=0$ the free energy per particle $f$ is the same as the internal energy $u$ since entropy is zero (or finite) for quantized systems in their ground states. For the quantized reference system the correlation function is (equal time correlations $\lambda=0$)
\begin{equation}
\rho^2 \tilde h_{ex}(k)=g \tilde S(0,k)-\rho.
\label{84}
\end{equation}
This leads to the exchange energy while the remaining part of the correlation function is the one that leads to the correlation energy
\begin{equation}
\rho^2 \tilde h_c(k)=g\tilde \Gamma(0,k)- g\tilde S(0,k)).
\label{85}
\end{equation}
With Eqs.~(\ref{63}) and (\ref{64}) for $u_c(k)$ which along with Eq.~(\ref{83a}) leads to $u_p=\lambda(\partial f/\partial \lambda)$, one altogether finds
\begin{equation}
p=\frac{2}{3}\rho f-\frac{1}{3}\rho\lambda\frac{\partial f}{\partial\lambda}.
\label{87}
\end{equation}
To see this  more clearly the free energy can be  split in three contributions
\begin{equation}
f=f_0+f_{ex}+f_c
\label{88}
\end{equation}
where for the reference system contribution, $\partial f_0/\partial\lambda$=0. For the exchange and correlation energies one with Eqs.~(\ref{84}) and(\ref{85}) has
\begin{equation}
\rho\lambda\frac{\partial f_{ex}}{\partial\lambda}=\frac{1}{2(2\pi)^3}\int d{\bf k}\, (g\tilde S(0,k)-\rho)\lambda\tilde\psi(k)
\label{89}
\end{equation}
\begin{equation}
\rho\lambda\frac{\partial f_c}{\partial\lambda}=\frac{1}{2(2\pi)^3}\frac{1}{2\pi}\int dK\int d{\bf k}\, g(\hat\Gamma(K,k)-\hat S(K,k))\lambda\tilde\psi(k).
\label{90}
\end{equation}
(Eq.~(\ref{90}) with variables of integration $K$ and ${\bf k}$ is the same as Eqs.~(\ref{61}) - (\ref{65}).) What remains is to show that the pressure (\ref{87}) also follows from the standard free energy route by differentiation with respect to density. One has the thermodynamic relation
\begin{equation}
p=\rho^2\frac{\partial f}{\partial \rho}.
\label{91}
\end{equation}
To show this we utilize the scaling form Eqs.~(\ref{40}) and (\ref{41}) for $f_c$ which also is valid for the whole free energy at $T=0$. With $\mu_f\sim1/r_s^2\sim\rho^{2/3}$ and $r_s\rightarrow \lambda r_s\sim\lambda/\rho^{1/3}$ we can write
\begin{equation}
f=\rho^{2/3} F(x), \quad x=\frac{\lambda}{\rho^{1/3}}.
\label{92}
\end{equation}
 Thus 
\begin{equation}
\rho\frac{\partial f}{\partial \rho}=\frac{2}{3}\rho^{2/3} F(x)+\rho^{2/3} F'(x)\left(-\frac{1}{3}x\right)=\frac{2}{3}f-\frac{1}{3}\lambda\frac{\partial f}{\partial\lambda}
\label{93}
\end{equation}
which inserted in Eq.~(\ref{91}) is the same as expression (\ref{87}). Thus result (\ref{87}) from the virial theorem is same as the one from the free energy route.

\end{document}